\documentclass[prl,twocolumn,preprintnumbers,amsmath,amssymb]{revtex4}
\usepackage{dcolumn}
\usepackage{bm,amsmath,amssymb}
\usepackage{graphicx}
\usepackage{bm}

\begin{document}

\title{Studies on phase transition and phase coexistence - traversing T and H using liquid helium}

\author{P. Chaddah and A. Banerjee}
\affiliation{UGC-DAE Consortium for Scientific Research\\University Campus, Khandwa Road\\
Indore-452017, M.P, India.}
\date{\today}
\begin{abstract}
Liquid helium is crucial to conveniently vary magnetic field (H) and temperature (T) giving access to a large region of (H, T) space; in stark contrast with the effort currently required to explore a comparable region of (P, T) space. The centenary of the liquefaction of helium is an appropriate occasion to talk on the utilization of our facilities for studies across first order phase transitions. We observed some interesting effects in variety of systems traversing (H, T) space across the magnetic first order phase transitions. We focus here on a few recent intriguing discoveries in the half-doped manganites related to the coexistence of equilibrium phase, with a glass-like phase having both structural and magnetic long-range order.
\end{abstract}
\maketitle

\section{Magnetic First Order Transition}
Phase transitions can be caused by varying one of the two (or more) independent control variables. The need to study the behavior in two-parameter space is clear for first order transitions if the Clausius-Clapeyron relation needs to be established. Superconducting magnets enable easy traversal of large regions in field and temperature. This talk on half-doped manganites shows results on supercooled and superheated states, and glass-like arrested states, which may provide new insights on metastabilities across first order transitions.

The Ehrenfest classification of phase transitions is based on observing a discontinuity, in some derivative of the free energy taken with respect to a control variable. For a first order transition, this discontinuity has to bear a specific relation with the slope of the phase transition line in the two parameter space. A lot of experimental effort goes into `establishing' a first order transition; the case of vortex solid-to-liquid transition is a widely researched recent problem where the two-parameter space explored was magnetic field (H) and temperature (T) \cite{zeldov, schilling}. This talk focuses on studies using facilities at our Consortium where liquid helium is imperative for varying both these parameters. 

Most of the materials of current interest, including the CMR manganites discussed here, are multicomponent systems whose properties become more interesting with substitutions \cite{pc1}. Frozen-in disorder is thus intrinsic to these materials and they would not exhibit sharp first order transitions; discontinuities would not be observed. The `rounded transitions', if first order, would show hysteresis related to supercooling and superheating. These metastable states would also be observed when the control parameter is other than temperature \cite{pc2}. It is interesting to establish that the transition is first order because discontinuities must then be occurring over the length scale of the correlation length, and this has obvious implications as one explores these compounds in nano-material form.

\section{Glass-like Arrest of First Order Transition}
A second metastability that could be observed if the rounded transition is first order corresponds to glass-like arrest of kinetics. This has the potential to \textbf{\emph{play havoc}} with theoretical efforts to understand a material because the state observed at temperatures below the closure of hysteresis may be a kinetically arrested state and not an equilibrium state \cite{ab1, ab2, ab3, pc3}. Resolving whether the observed state is an equilibrium state becomes possible only by traversing the two-parameter space (of H and T, or of P and T) and we shall amplify on this with our data on CMR manganites. Liquid helium is crucial to our ability to vary H over the range of -14 Tesla to 14 Tesla. It is also crucial to our ability to vary T from 2K to ambient, and the centenary of liquefaction of helium is an appropriate occasion to talk of this work. We can span the space of two thermodynamic parameters H and T, and study the stability of different phases. Since H, unlike P, is transmitted without a medium, this is experimentally easier than spanning P and T. Transition temperature T$_C$ would vary with the second thermodynamic parameter for a first order transition; one finds a much larger variation of T$_C$ with the experimentally available range of H (in some magnetic/superconducting systems) than one finds with the experimentally available range of P. In this sense, we can today span a much larger region of (H,T) space, than of (P,T) space.

\emph{Manganites provide an excellent platform for such studies; no other known family of materials provides the versatility detailed below.} With slight change of composition, the ground state can be changed from ferromagnetic-metal (FM-M) to antiferromagnetic insulator (AF-I). Because of this, the slope of the first order transition line (in H-T space) changes sign within the same family of materials \cite{ab1, ab2}. Again, glass-like kinetic arrest occurs at a temperature T$_K$ which is a function of H. This dependence of T$_K$ on H also changes sign within the manganite family as one goes from FM-M ground state to AF-I ground state \cite{ab1}. Finally, since the conductivity changes drastically along with magnetic order, this family of CMR manganites has an inherent advantage over studies on other metamagnetic materials (with no accompanying metal to insulator transition). A decrease in global magnetization of the sample can be interpreted as either reduction of moment in FM-M phase, or as part-transformation of FM-M to AF-I. A simultaneous measurement of conductivity provides a clear choice between the two alternatives because of the orders of magnitude resistivity changes associated with the metal (M) to insulator (I) transition in the latter case. \emph{This is a special for half-doped manganites, and obviates the necessity of mesoscopic measurements to establish phase coexistence. }
 
We show in fig 1 magnetization (M vs T in H=1Tesla) and resistivity (R vs T in H=0Tesla) measurements for the half-doped manganite La$_{0.5}$Ca$_{0.5}$MnO$_3$ (LCMO) \cite{ab3, pc3}.
\begin{figure}[h]
	\centering
		\includegraphics{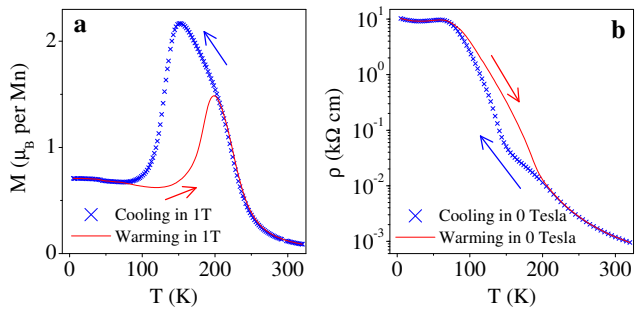}
	\caption{First order ferromagnetic-metallic to antiferromagnetic-insulating transition in La$_{0.5}$Ca$_{0.5}$MnO$_3$. (a) Magnetization while heating and cooling in 1T field. (b) Resistivity while heating and cooling in zero field.}
	\label{fig:Fig1}
\end{figure}
These bring out that, with decreasing T, the sample undergoes a ferromagnetic-metallic to antiferromagnetic-insulating transition. We note a large thermal hysteresis corresponding to the metastable supercooled and superheated states.

We show in fig 2 magnetization (M vs T in various H from 3Tesla to 6Tesla) and resistivity (R vs T in various H from 3Tesla to 4Tesla) measurements for the half-doped manganite Pr${_{0.5}}$Ca$_{0.5}$MnO${_3}$ (PCMO), with 2.5\% Al substitution on Mn site (PCMAO) \cite{ab2}. These bring out that, with decreasing T, this sample undergoes an antiferromagnetic-insulating to ferromagnetic-metallic transition. We again note a large thermal hysteresis corresponding to the metastable supercooled and superheated states. We also note that the reversible values of M and R, obtained below the closure of hysteresis, show that the AF-I to FM-M transition is not completed even at the lowest temperature. This glass-like arrest of the high-temperature phase corresponds to a metastability that is very different from supercooling \cite{ab3, pc3}.

\begin{figure}[h]
	\centering
		\includegraphics{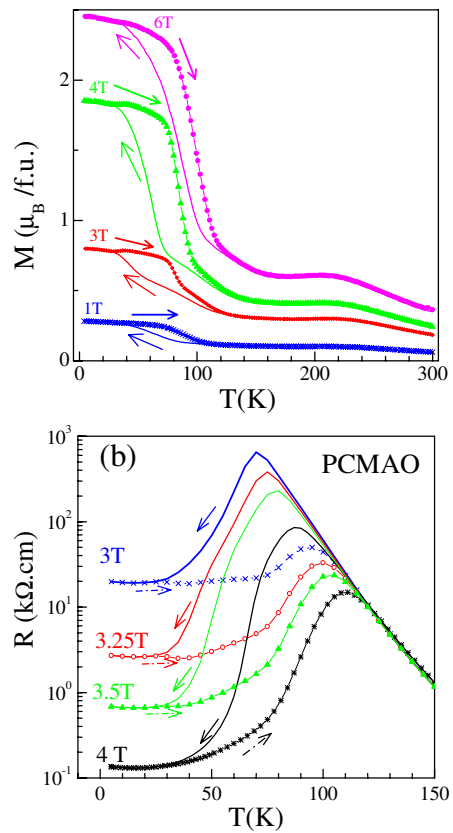}
	\caption{Thermal hysteresis in magnetization and resistivity in same measurement field for Pr${_{0.5}}$Ca$_{0.5}$Mn$_{0.975}$Al$_{0.025}$O${_3}$ sample showing first order AF-I to FM-M transition. The measurements are repeated for different measurement fields. (a) shows magnetization while cooling and then heating in the same field. The measurements field ranges between 1-6 T. Steep rise in magnetization around 100 K indicate AF to FM transition. (b) shows resistivity while cooling and then heating in same field. The measurement fields ranges between 3-4 T.  Sharp fall in resistivity  around 100 K indicate insulator to metal transition.}
	\label{fig:Fig2}
\end{figure}

\section{Glass formation: Rapid vs. Slow Cooling}
A glass is viewed as a liquid with time held still \cite{Braw}, a liquid in which the molecules have suddenly stopped moving at some instant of time. The motion is frozen by reaching a low temperature ($<$T$_g$), and the sites are frozen by reaching this T on a time scale that is very short compared to the time required for a molecule to adjust its position. This is the philosophy underlying the splat-cooling technique used to form metallic glasses. It is known, however, that the cooling rate required for glass formation depends on the ratio of T$_g$ to the thermodynamic freezing point \cite{greer}. 

We have been, however, comparing T$_g$ with the spinodal limit T* for supercooling, and have argued that slow cooling can result in glass formation if T$_g$ $>$ T*, whereas rapid cooldown is essential if T$_g$ $<$ T* \cite{pc1, pc3}. We have also generalized the term glass to include any high-T phase (its higher entropy implies higher disorder, even if it is not of the structural kind) that exists at low-T; its decay rate decreasing with lowering T \cite{ab3, pc3}. In this category of generic glass, half-doped manganites provide first order transitions where T$_C$ (and T*) is tuned heavily by varying H. This opens up the possibility that T$_g$ $>$ T* at some H, and T$_g$ $<$ T* at some other H, in the same material \cite{pc1}. Slow cooling at the former field value will yield a glass, which will devirtify on slow warming in the latter field value. We have accordingly exploited the new measurement protocol CHUF (Cooling and Heating in Unequal Fields) to access, in a controlled way, both glass formation and glass devitrification. We now provide data as we explore this new possibility, and shall discuss some physics results with general applicability in the last section. In all the measurements reported here, temperature was varied at 1.5K/min, which is our version of slow cooling/warming.

\section {Results on Half-doped Manganites}
We show in fig.3 magnetization measurements in Al. doped PCMO (PCMAO).
\begin{figure}[h]
	\centering
		\includegraphics{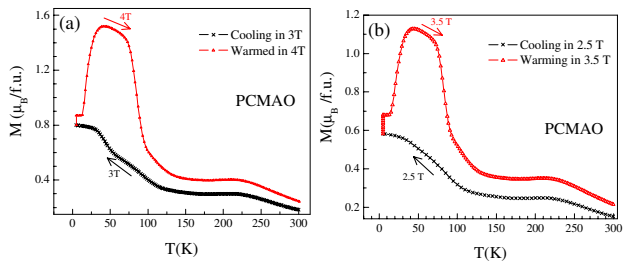}
	\caption{Depicts devitrification of arrested AF-I phase through magnetization measurements. (a) shows magnetization while cooling in 3 T. Then at 5 K the field is isothermally increased to 4 T and magnetization is measured while warming in 4T. The sharp increase in magnetization above 15 K shows the devitrification of the remaining Af-I phase. (b) shows similar devitrification while warming in 3.5 T after cooling in 2.5 T.}
	\label{fig:Fig3}
\end{figure}  This material is close to half-doping but the small amount (2.5\%) of Al on Mn site changes the low temperature ground straight to FM-M \cite{ab2}. On cooling in H=3T (fig.3a) or in H=2.5T (fig.3b), we find an anti-ferromagnetic to ferromagnetic transition initiated at about 100K and continuing till about 30K. If the sample is then warmed in a higher field (the field is changed isothermally at 5K), we find a further anti-ferromagnetic to ferromagnetic transition between 20K and 35K. We explain the increase in magnetization on lowering T as due to the first order AF-I to FM-M transition that undergoes a glass-like arrest at about 35K, before it can be completed. In a higher field applied during warming, the glass-like arrest temperature has become lower while T* has become higher, and devitrification of the arrested AF-I state is observed \cite{ab4}.

In fig.4 the same physics is shown through resistivity measurements.
\begin{figure}[h]
	\centering
		\includegraphics{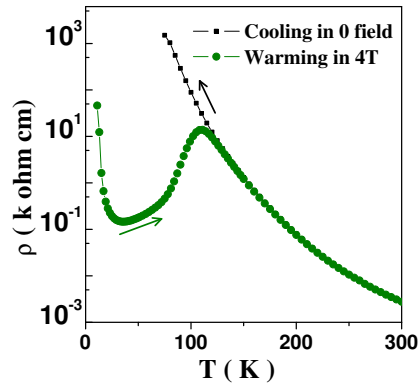}
	\caption{Resistivity while cooling in zero field increases monotonically with decrease in temperature and exceeds the measurement range of the instrument below 75 K. After cooling to 5 K, the field is isothermally changed to 4 T and resistivity is measured while warming. The rapid fall with increase in temperature indicates devitrification of the glassy AF-I phase.}
	\label{fig:Fig4}
\end{figure}
 The transformation to the FM-M state is not seen on cooling in H=0, and the insulating (AF-I) state is arrested.  Since T$_g$ falls in higher field we observe devitrification to the metallic state (FM-M) on warming in the higher field of 4 Tesla \cite{ab2}. The same behavior of devitrification from AF-I to FM-M is shown in fig.5 for La-Pr-Ca-Mn-O,\begin{figure}[h]
	\centering
		\includegraphics{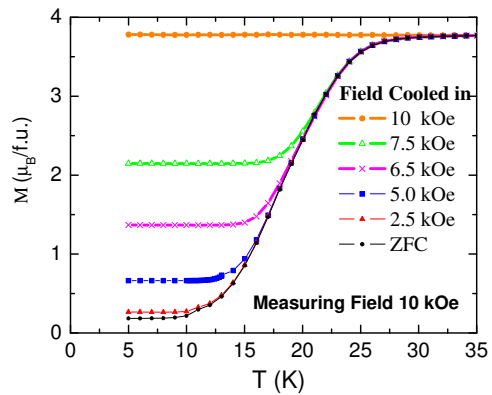}
	\caption{Magnetization of La-Pr-Ca-Mn-O in 1 T while warming after cooling to 5 K in different field. The rapid increase indicates devitrification of the arrested AF-I phase to equilibrium FM-M phase. }
	\label{fig:Fig5}
\end{figure} a manganite that is far from half-doping \cite{kranti}.
Here the sample is warmed in a field (H=1T) higher than the various cooling fields used, and devitrification is seen below 30K.

We now consider the half-doped manganite LCMO whose ground state is AF-I.  Here glass-like arrest of the high temperature FM-M phase is observed with cooling in a high field, and devitrification to the AF-I phase is seen warming in lower field.  This is brought out through magnetization (Figs.6a \& 6b) and resistivity (Figs. 6c \& 6d) measurements \cite{ab3, pc3}.

\begin{figure}[h]
	\centering
		\includegraphics{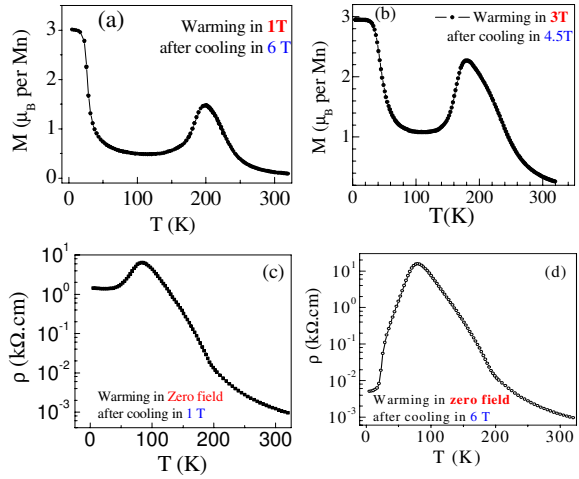}
	\caption{Devitrification of arrested FM-M state in La$_{0.5}$Ca$_{0.5}$MnO$_3$ shown through magnetization and resistivity measurements while warming in higher fields than the cooling fields. (a) Magnetization in 1 T after cooling in 6 T and isothermally changing the field to measurement field of 1 T at 5 K. (b) similar to (a), measured in 3 T after cooling in 4.5 T. (c) Resistivity while warming in zero field after cooling in 1 T and isothermally changing the field to measurement field of 0 T at 5 K. (d) similar to (c), measured in zero field after cooling in 6 T. }
	\label{fig:Fig6}
\end{figure}
Devitrification can also be observed on isothermal variation of H. \begin{figure}[h]
	\centering
		\includegraphics{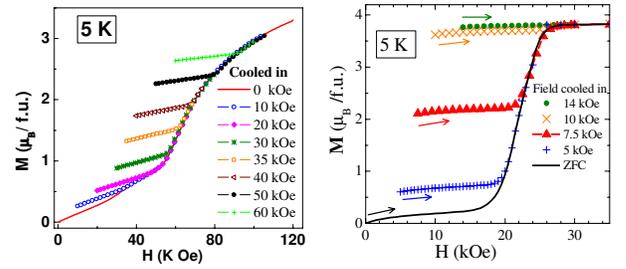}
	\caption{Depicts devitrification of arrested AF-I phase during isothermal field variation. (a) Pr${_{0.5}}$Ca$_{0.5}$Mn$_{0.975}$Al$_{0.025}$O${_3}$ sample is cooled from room temperature to 5 K in different fields. Then magnetization is measured at 5 K while increasing the field from the value of the respective cooling fields. (b) devitrification of arrested AF-I phase of La-Pr-Ca-Mn-O system during isothermal field variation. The sample is cooled from room temperature to 5 K in different fields. Then magnetization is measured at 5 K while increasing the field from the value of the respective cooling fields. It may be noted that for both the cases the sharp change in magnetization shifts to higher field values for higher cooling fields.}
	\label{fig:Fig7}
\end{figure}
When the arrested state is AF-I then devitrification is observed on raising H. We show results through magnetization measurements for PCMAO (Fig. 7a) and for La-Pr-Ca-Mn-O (Fig. 7b) \cite{ab2, kranti}.  If the arrested state is FM-M as in pure half-doped manganites, then devitrification will be observed on lowering H. \begin{figure}[h]
	\centering
		\includegraphics{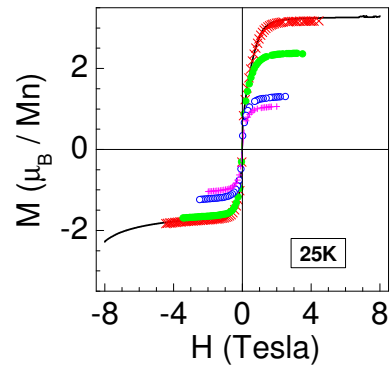}
	\caption{Evidence of kinetically arrested glassy FM-M phase fractions in La$_{0.5}$Ca$_{0.5}$MnO$_3$ and their devitrification during reduction of field. Each time the sample is cooled from 320 K to 25 in different field H. Then the magnetization is measured while isothermally cycling the field at 25 K from H to -H. Cooling in different H renders different amount of FM-M phase at 25 K. During the field reduction from H to 0, they show distinct magnetization because of different amount of frozen FM-M phase. However, as the field is reduced to zero a part of these arrested FM-M fractions devitrifies to equilibrium AF-I phase as is evident in lower magnetization values in the negative field cycling. The devitrification is more for the higher cooling fields, which is to do with the anticorrelation between supercooling and kinetic arrest (glass transition) temperatures of different regions.}
	\label{fig:Fig8}
\end{figure}  In this case devitrification is seen on lowering H at 25K (Fig. 8) \cite{ab3}.

In other studies on manganites we have also focused on the broad first order transition which corresponds to different regions of the sample (of length scale corresponding to the correlation length) having different values of T$_C$, and also of T$_g$.  Through macroscopic measurements following non-conventional paths in (H,T) space, we have shown in various manganite samples that region which have lower T$_C$ (or T*) have a higher T$_g$ \cite{ab3, pc3, kranti, rawat1}. This anti-correlation between the effect of disorder on T$_g$ and T* has also been observed in some non-manganite samples showing first order magnetic transitions \cite{roy, rawat2}. Does this correspond to the confusion principle enunciated for metallic glasses in the context of the T$_g$? \cite{greer}   Similar studies with pressure (instead of H) being the second control parameter would be needed to check whether the anti-correlation being observed here is a general principle extending to all kinds of glasses.

\section {Conclusions of general applicability}
In our recent studies on LCMO and on PCMAO, we have found that if we chose a cooling field so as to have a larger fraction of glass at low temperature, then subsequent devitrification and recrystallization results in a larger faction of equilibrium phase \cite{ab3, ab4}.  The question of this strong initial state dependence being applicable to all the structural glasses is an interesting one. 

\section{Acknowledgement}
DST Government of India is acknowledged for funding the 14 Tesla PPMS-VSM.

\end{document}